# A Quiet Quantum Revolution in Earth's Deep Interior

***Renata Wentzcovitch*** (*rmw2150@columbia.edu*) *Departments of Earth and Environmental Science and Applied Physics and Applied Mathematics, Lamont-Doherty Earth Observatory, Columbia University, New York, NY (US);* ***Laura Cobden*** *(L.J.Cobden@uu.nl) Utrecht University, Utrecht (NL);* ***Christine Houser*** *(chouser@elsi.jp) Earth-Life Science Institute, Tokyo (JP),* ***Grace Shephard*** *(grace.shephard@anu.edu.au) Research School of Earth Sciences, Australian National University, Camberra (AU);* ***Jingyi Zhuang*** *(jz2907@columbia.edu), Departments of Earth and Environmental Science, Lamont-Doherty Earth Observatory, Columbia University, New York, NY (US).*

***Abstract*** *- The Earth's lower mantle hosts a subtle but pervasive quantum phenomenon: the pressure-induced spin crossover of iron in its dominant minerals, bridgmanite and ferropericlase. In this transition, iron ions gradually shift from high-spin to low-spin electronic states without structural change, altering their volume, compressibility, and elastic properties. Although long recognized experimentally and theoretically, its geophysical significance has only recently become clear through the integration of mineral physics and three-dimensional seismic imaging. The spin crossover reduces bulk modulus and P-wave velocities while leaving S-wave speeds largely unaffected, producing a distinctive decoupling between P- and S-wave anomalies. This signature is now observed in global tomography and reconciles seismic observations with realistic mantle temperatures and compositions. Rather than forming a sharp boundary, the crossover extends across most of the lower mantle, acting as a diffuse yet essential control on seismic structure. This work highlights how quantum-scale electronic transitions influence planetary-scale dynamics and interpretations of Earth's deep interior.*

At depths below 1,000 kilometers inside the Earth, pressures exceed one million atmospheres, and temperatures are hotter than lava at the surface. There, the rocks of Earth's mantle behave in ways that defy intuition. For decades, seismologists have interpreted the structure of this region through three central ideas: phase transitions that rearrange crystal structures, temperature differences associated with mantle convection, and compositional variations that reflect billions of years of recycling and differentiation. These processes have shaped the standard explanation for why seismic waves travel at different speeds throughout the Earth. Yet researchers have uncovered a fourth contributor, one neither structural, thermal, nor chemical. Instead, it arises from the quantum behavior of electrons in iron ions.

Deep in the lower mantle, iron in the two dominant minerals, bridgmanite and ferropericlase, undergoes a pressure-driven rearrangement of its electronic structure known as a spin transition. In this process, the d-electrons of an iron ion change from a "high-spin" configuration to a "low-spin" one. No bonds are broken, no atoms shift position, and the crystal symmetry remains unchanged. But iron shrinks and its magnetic moment collapses. High-pressure experiments by Badro et al. [1], Lin et al. [2], Komabayashi et al. [3], and others revealed this electronic transformation in the laboratory two decades ago.

At the time, the implications for Earth were considered important but speculative. How could a phenomenon happening inside individual ions matter at the scale of continents and convection flows? The answer has emerged only gradually: magnetic moment, or spin, is an intrinsic property in single ions, and the populations of high- and low-spin irons change continuously with pressure and temperature [4], subtly altering the compressibility of the host phase [5]. For this reason, this phenomenon is referred to as the iron spin crossover (ISC). Ferropericlase ((Mg, Fe)O)) makes up around 20% of the lower mantle volume, but has the highest iron content and therefore the ISC is most detectable in this phase. Meanwhile, the ISC in bridgmanite ((Mg,Fe)(Si,Fe)$O_3$), which constitutes up to 80% of the lower mantle, might not have a significant effect since it requires $Fe^{3+}$ to replace silicon in the perovskite octahedral site. This is not something that happens easily because aluminum is the favored substitutional ion and $Fe^{3+}$ is expected to be much less abundant than $Fe^{2+}$. Because compressional seismic waves (P-waves) respond to changes in compressibility, the ISC leaves a distinctive imprint on P-wave speeds, while shear-wave speeds (S-wave or Vs) change much less [6].

This difference is hidden in plain sight. Only when mineral physics and 3-D seismic imaging were brought together did the consequences become clear.

### The Physics of a Spin Crossover

The ISC is a strictly quantum phenomenon. In a high-spin state, electrons occupy separate orbitals with parallel spins, resulting in a larger ion with a large magnetic moment. In a low-spin state, electrons pair within lower-energy orbitals, reducing both spin and radius. The competition between crystal-field splitting ($E_c$, favoring low spin) and Hund's rule coupling ($E_X$, favoring high spin) determines which state is stable (Fig. 1A). Temperature determines their relative abundances (Fig. 1B).

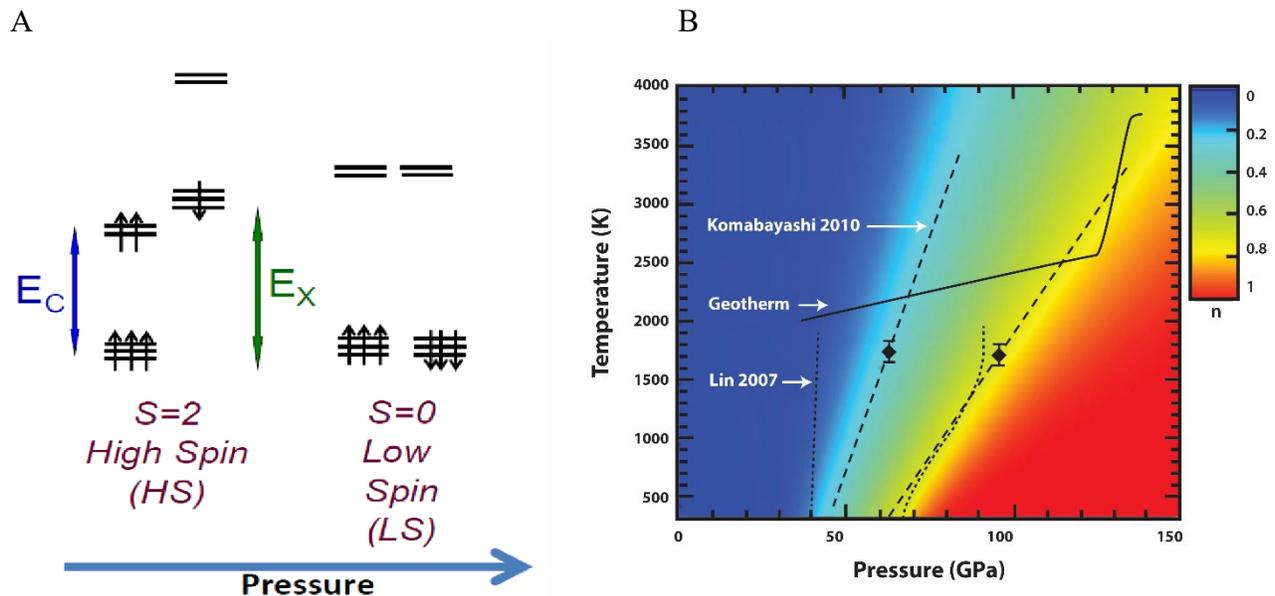

**Figure 1.** (A) *Iron in the octahedral site of lower mantle minerals changes from a high-spin to a low-spin electronic state with increasing pressure/depth. (B) It results in a broad transition in the pressure range, with continuous changes in high-spin and low-spin iron populations, which increase with temperature [9].*

At ambient pressure, the crystal field in both ferropericlase and bridgmanite is too weak to force electron pairing, so iron remains in the high-spin state. But increasing pressure strengthens the crystal field, stabilizing the low-spin configuration. Laboratory measurements showed that iron's spin state starts changing at depths near 1,000 kilometers and continues well past 2,000 kilometers, covering most of the lower mantle [7].

This wide, mixed-spin region, confirmed by experiments and *ab initio* calculations [7-9], indicates that the ISC does not create a sharp seismic discontinuity. Instead, it produces a broad mixed-spin region in which high- and low-spin states coexist (Fig. 1B), each with its characteristic ionic radius and local strains.

Those structural subtleties and changes in the population of these iron types under pressure alter their responses to compression.

## From Quantum Physics to Seismic Velocities

The gradual pressure-induced ISC reduces the bulk modulus (Fig. 2), i.e., the resistance to compression, while the shear modulus, the resistance to shape deformation, gently increases [6]. Because S-waves depend on the shear modulus and P-waves depend on both moduli, the ISC reduces Vp, while Vs increases gently in most of the lower mantle.

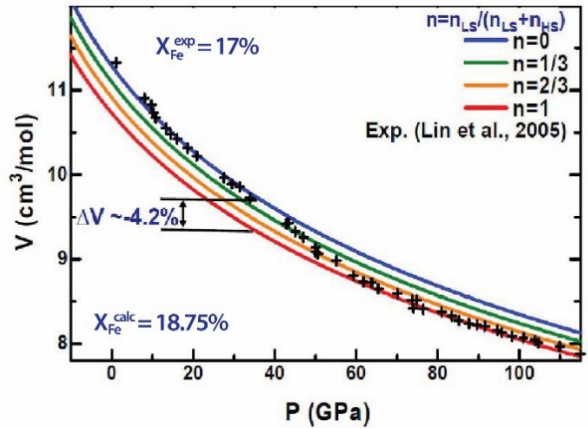

**Figure 2.** *The ISC reduces ferropericlase's resistance to compression [5], lowering P-wave speeds across a broad depth interval while leaving S-wave speeds nearly unchanged. This effect produces characteristic P–S wave patterns in three-dimensional tomography*

A breakthrough came when advanced *ab initio* simulations by Zhuang and Wentzcovitch [9] revealed that the elastic signature of the ISC is very diffuse, spanning most of the lower mantle, broadly reducing P-velocity speeds, rather than abrupt decreases in velocity. Because the ISC is temperature-dependent, these broad variations are even less obvious when 3D seismic wave speeds are spherically averaged, making them less apparent in 1D models (Fig. 3).

Yet these updated models also showed that the ISC-induced reduction in P-wave speed is unavoidable. Without it, mantle temperatures inferred from seismic velocities would need to be unrealistically low, and compositions would have to be extreme [10]. In this sense, the ISC acts as a background correction: essential but invisible unless explicitly accounted for (Fig. 3).

This insight set the stage for a new way of looking at seismic data, not through 1-D radial averages, but through the three-dimensional relationships between P- and S-wave speeds.

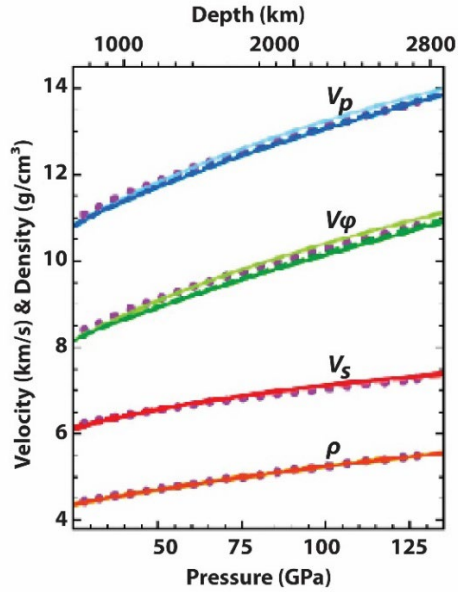

**Figure 3.** *Predicted velocities of pyrolite (75 vol% bridgmanite, 18 vol% ferropericlase, 7 vol% davemaoite, with $X_{Fe}$ = 0.10) compared to PREM (light/dark colors are properties in the assemblage without/with ISC in ferropericlase. $V_P = \sqrt{(K + {}^4/_3 \mu)/\rho}$, $V_S = \sqrt{\mu/\rho}$, and $V_\varphi = \sqrt{K/\rho}$, are compressional, shear, and bulk wave speeds and depend on bulk (K) and shear (μ) moduli and density (ρ). Solid/dashed lines indicate the presence/absence of iron partitioning effects. The mixed-spin region of ferropericlase spans most of the lower mantle, producing subtle, widespread effects on seismic velocities [9]. This diffuse depth range explains why one-dimensional global models do not exhibit obvious anomalies [9].*

## A Seismological Signature Noticeable in 3D

Where the ISC becomes strikingly evident is in the difference between P-wave and S-wave speed anomalies induced by temperature variations (Fig. 4). Because $V_P$ is affected differently than $V_S$, regions containing ferropericlase should show muted P-wave anomalies relative to their S-wave anomalies.

Slabs of subducted oceanic lithosphere, for example, have faster wave speeds than the ambient mantle because they are colder. In this case, thermodynamic models suggest the spin crossover should dampen their P-wave anomalies but not S-wave anomalies. This produces a diagnostic pattern: decorrelation of P–S wave speeds.

This effect was recognized by Wu and Wentzcovitch [8], but the first global-scale evidence of this appeared in Shephard et al. (2021) [11], who used "vote maps", a method that highlights consistent features across many independent tomography models. These maps showed that in the lower mantle, P- and S-wave structures diverge precisely where ferropericlase is expected to be in a mixed-spin state.

More recently, Cobden et al. [10] used full-waveform tomography to obtain absolute P- and S-wave speeds. Their results show that reproducing observed velocities with realistic temperatures and compositions requires inclusion of the iron spin crossover; without it, the mid-mantle would have to be unrealistically cold and strongly depleted in silicon.

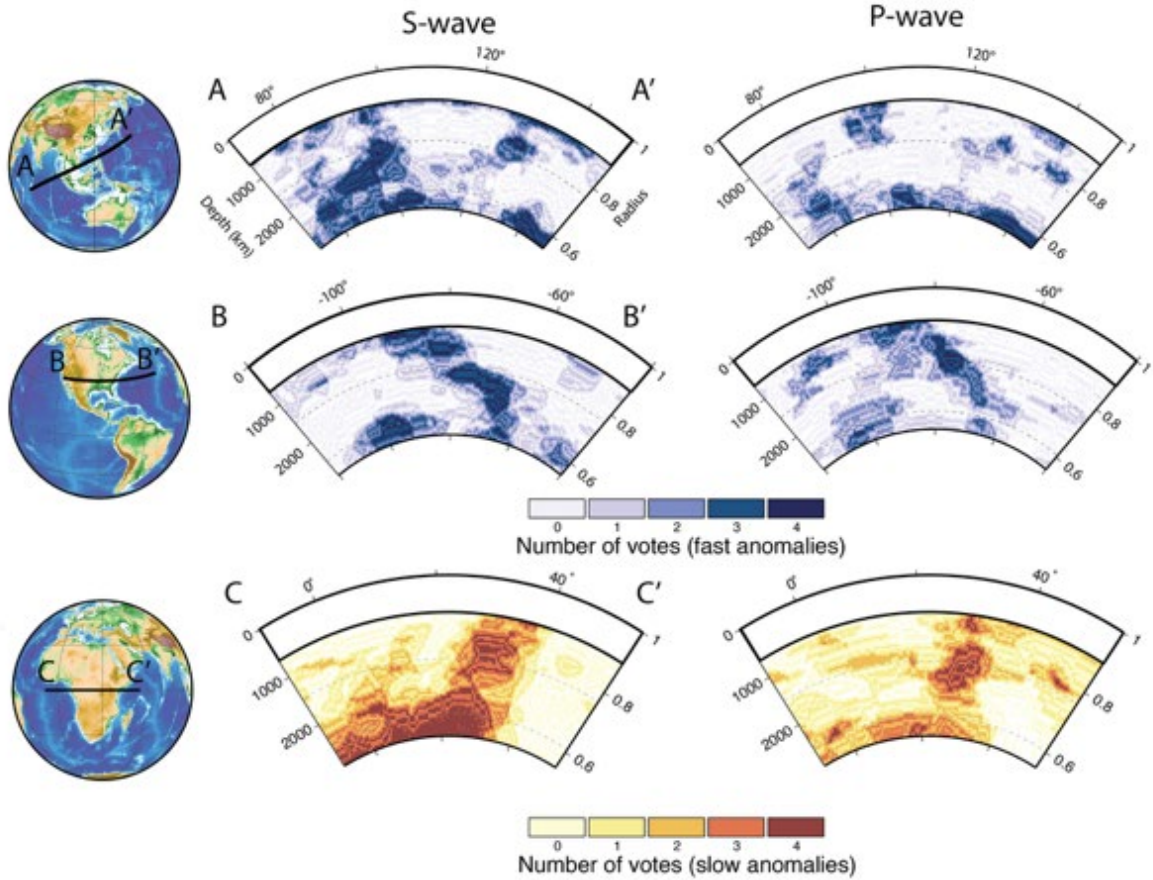

**Figure 4.** *Comparisons of global P- and S-wave tomography using the "vote map" method show the predicted signature of the iron spin crossover: seismically fast, cold, ferropericlase-rich slabs (a and b) and seismically slow, hot, plume areas (c) exhibit weaker P-wave anomalies but stronger S-wave anomalies. This P–S decorrelation is consistent with mineral-physics predictions and is observed in both the vote map [11] and the full-waveform models [10].*

Although the influence of the iron spin crossover on the seismological parameter $R_{s/p}$ had been previously recognized [8], Zhuang and Wentzcovitch [9] demonstrated that a fully non-ideal solid-solution treatment of the ISC in ferropericlase quantitatively reconciles mineral-physics predictions with seismological observations.

Together, these findings form a consistent narrative: the ISC is not expressed as a sharp or diffuse "layer," but through three-dimensional contrasts between P- and S-wave structures.

## A Mantle Quietly Reshaped by the Iron Spin Crossover

First, it must be recognized that the iron spin crossover (ISC) reshapes interpretations of deep-mantle seismic structure. Seismic heterogeneity reflects not only temperature and composition but also variations in spin state abundances. Temperature-dependent changes in high- and low-spin iron populations directly influence wave speeds. Second, including ISC effects yields more realistic mantle temperatures in full-waveform tomography interpretations. Without it, models require unrealistically cold or compositionally extreme regions. Third, mixed-spin ferropericlase densifies more rapidly with pressure, invigorating mantle convection [12], and may reduce viscosity, potentially influencing slab sinking, stagnation, and plume morphology.

## Why This Matters for Geophysics

The ISC illustrates how quantum-scale processes shape planetary-scale behavior. Rather than producing sharp seismic boundaries, it introduces subtle, pervasive effects on wave speeds and their relationships in 3D. It provides seismologists with a framework for reconciling temperature, composition, and velocity anomalies; highlights for mineral physicists the importance of electronic structure; and suggests to geodynamicists that electron spin may influence buoyancy, viscosity, and flow. More broadly, it shows that key processes in Earth's interior can remain hidden until multiple disciplines are integrated.

As seismic imaging improves and mineral-physics models become more accurate, the role of the ISC in shaping lower mantle structure and dynamics will become clearer. Its influence on LLSVPs, the D″ discontinuity and layer, and slab stagnation zones remains an active area of research. What is already clear is that the ISC is a global, depth-spanning process essential to understanding Earth's deep interior.